\documentclass[prl,twocolumn,superscriptaddress]{revtex4-1}

\usepackage{amsmath}
\usepackage{amssymb}
\usepackage{amsthm}
\usepackage{amsfonts}
\usepackage{listings}
\usepackage{enumerate}
\usepackage{latexsym}
\usepackage{color}
 \usepackage{multirow}
\usepackage{physics}
\usepackage{bm}
\usepackage{hyperref}
\hypersetup{
 pdfnewwindow=true, colorlinks=true,
 linkcolor=blue, anchorcolor=blue,
 citecolor=blue, filecolor=blue,
 menucolor=blue, urlcolor=blue}

\usepackage{psfrag}

\usepackage{bm}
\usepackage{graphicx}
\usepackage{subfigure}

\RequirePackage[normalem]{ulem} 
\RequirePackage{color}\definecolor{RED}{rgb}{1,0,0}\definecolor{BLUE}{rgb}{0,0,1} 

\newcommand{\bk}{{\vb* k}}

\newcommand{\bB}{{\vb* B}}
\newcommand{\bA}{{\vb* A}}
\newcommand{\be}{{\vb* e}}

\newcommand{\pref}[1]{(\ref{#1})}

\def\ie{{\it i.e.},\ }

\input{epsf}

\begin{document}

\tolerance 10000

\newcommand{\vk}{{\bf k}}

\draft

\title{Majorana Zero Modes and Topological Nature in Bi$_{2}$Ta$_3$S$_6$-family Superconductors}

\author{Yue~Xie}
\thanks{These authors contributed equally to this work.}
\affiliation{Beijing National Laboratory for Condensed Matter Physics,
and Institute of Physics, Chinese Academy of Sciences, Beijing 100190, China}
\affiliation{University of Chinese Academy of Sciences, Beijing 100049, China}

\author{Zhilong~Yang}
\thanks{These authors contributed equally to this work.}
\affiliation{School of Mathematics and Physics, University of Science and Technology Beijing,
Beijing 100083, China
}

\author{Ruihan~Zhang}
\affiliation{Beijing National Laboratory for Condensed Matter Physics,
and Institute of Physics, Chinese Academy of Sciences, Beijing 100190, China}
\affiliation{University of Chinese Academy of Sciences, Beijing 100049, China}

\author{Sheng~Zhang}
\affiliation{Beijing National Laboratory for Condensed Matter Physics,
and Institute of Physics, Chinese Academy of Sciences, Beijing 100190, China}
\affiliation{University of Chinese Academy of Sciences, Beijing 100049, China}

\author{Quansheng~Wu}
\affiliation{Beijing National Laboratory for Condensed Matter Physics,
and Institute of Physics, Chinese Academy of Sciences, Beijing 100190, China}
\affiliation{Condensed Matter Physics Data Center, Chinese Academy of Sciences, Beijing 100190, China}

\author{Gang Wang}
\affiliation{Beijing National Laboratory for Condensed Matter Physics,
and Institute of Physics, Chinese Academy of Sciences, Beijing 100190, China}

\author{Hongming~Weng}
\affiliation{Beijing National Laboratory for Condensed Matter Physics,
and Institute of Physics, Chinese Academy of Sciences, Beijing 100190, China}
\affiliation{Condensed Matter Physics Data Center, Chinese Academy of Sciences, Beijing 100190, China}

\author{Zhong Fang}
\affiliation{Beijing National Laboratory for Condensed Matter Physics,
and Institute of Physics, Chinese Academy of Sciences, Beijing 100190, China}
\affiliation{Condensed Matter Physics Data Center, Chinese Academy of Sciences, Beijing 100190, China}

\author{Xi~Dai}
\email{daix@ust.hk}
\affiliation{Department of Physics, Hong Kong University of Science and Technology, Hong Kong 999077, China}

\author{Zhijun~Wang}
\email{wzj@iphy.ac.cn}
\affiliation{Beijing National Laboratory for Condensed Matter Physics,
and Institute of Physics, Chinese Academy of Sciences, Beijing 100190, China}
\affiliation{Condensed Matter Physics Data Center, Chinese Academy of Sciences, Beijing 100190, China}

\date{\today}

\begin{abstract}
In this work, we report that Bi$_{2}$Ta$_3$S$_6$-family superconductors exhibit nontrivial band topology. They possess a natural quantum-well structure consisting of alternating stacks of TaS$_2$ and honeycomb Bi layers, which contribute superconducting and topological properties, respectively. Symmetry-based indicators $(\mathbb{Z}_4;\mathbb{Z}_{2}\mathbb{Z}_{2}\mathbb{Z}_{2})=(2;000)$ reveal that the topological nature arises entirely from the Bi layers, which belong to a quantum spin Hall phase characterized by a $p_x$–$p_y$ model on a honeycomb lattice. The topological zigzag (ZZ) and armchair (AC)  edge states are obtained. Using VASP2KP, the in-plane $g$ factors of these topological edge states are computed from the \emph{ab~initio} calculations: $g_{x/y}^{\mathrm{ZZ}}=2.07/1.60$ and $g_{x/y}^{\mathrm{AC}}=0.50/0.06$. The strong anisotropy of the edge-state $g$ factors allows us to explore Majorana zero modes in the Bi monolayer on a superconductor, which can be obtained by exfoliation or molecular beam epitaxy. The relaxed structures of the Bi$_2$Ta$_3$Se$_6$, Bi$_2$Nb$_3$S$_6$ and Bi$_2$Nb$_3$Se$_6$ are obtained. Their superconducting transition temperature $T_c$ are estimated based on the electron-phonon coupling and the McMillan formula. Furthermore, using the experimental superconducting gap $\Delta$ and the computed $g$ factors, we obtain the phase diagram, which shows that the in-plane field $B_y>2.62~\mathrm{T}$ can generate corner Majorana zero modes in the Bi monolayer of the superconductor Bi$_{2}$Ta$_3$S$_6$.
A similar paradigm also applies to the Bi$_{2}$Ta$_3$S$_6$ bulk with the emergence of Majorana hinge states.
These natural quantum-well superconductors therefore offer ideal platforms for exploring topological superconductivity and Majorana zero modes.    
\end{abstract}

\maketitle


\paragraph*{Introduction.---}
Majorana zero modes (MZMs) in topological superconductors (TSCs)~\cite{review1,review2,review3,review4} have attracted great interest over the past two decades because of their non-Abelian exchange statistics and potential applications in topological quantum computation~\cite{review5,review6}. A $p$-wave superconductor is intrinsically an topological superconductor~\cite{1DpwaveSC,2DpwaveSC}. However, it is so rare and no $p$-wave superconductors have been found in experiments yet.
To realize TSCs and MZMs, an alternative route is to introduce spin-singlet superconductivity into systems with nontrivial band topology~\cite{review_Alicea,interface-SC-on3DTI} via the proximity effect, either extrinsically or intrinsically~\cite{proximity-BiSe,proximity-BiSe-2,self-proximity,self-proximity-FeSe}. The boundaries of a topological insulator decorated with superconductivity become effective $p$-wave superconductors~\cite{TSC_FuKane,Alicea}, supporting vortex MZMs on 2D surfaces~\cite{vortexMZM,Cn-MZM,anomalous-vortex}, Majorana chiral/helical modes along 1D edges~\cite{chiralMZM-Qi,chiralMZM-Wang,chiralMZM-Nagaosa,helicalMZM} or MZMs at 0D corners~\cite{MKP-eswave,MKP-dwave,gapcompete,cornerMZM-prl,cornerMZM-prb}. Following this guiding ideology of superconductivity-band topology hybridization, candidates of TSCs have proliferated in the community, including 
iron-based superconductors FeSe/FeTe family~\cite{FeSe_Wang,FeTeSe-Wu,vMZM-Xu,exp-FeSeband,vMZM-weakTI,PdBiTe,Zhang-1,Zhang-2,Zhang-3,compete_vortex,dislocation-MZM}, superconducting topological crystalline insulators such as SnTe~\cite{Mz-MZM,SnTe-exp}, and very recent unconventional materials such as 1$T$-PtSe$_2$ and Nb$_3$Br$_8$~\cite{1T-PtSe2-TSC,Nb3Br8-TSC}.
However, among these superconductivity-band topology combining systems, most of the reproducible experiments have been limited to MZMs bound to the vortex flux quantum~\cite{vMZM-BiSe-1,vMZM-BiSe-2,vMZM-FeTeSe-1,vMZM-FeTeSe-2,vMZM-FeTeSe-3,vMZM-FeTeSe-4,vMZM-FeTeSe-5,Skymion-MZM} or on possible Majorana states propagating along 1D domains~\cite{chiralMZM-1,chiralMZM-2}. Other kinds of Majorana modes, such as corner MZMs predicted in the corners of quantum spin Hall insulators (QSHIs) with anisotropic Zeeman effects~\cite{cornerMZM-prl}, remain out of evidence due to the lack of realistic material candidates.

Recently, a layered compound Bi$_2$Ta$_3$S$_6$ was reported to be a superconductor with a superconducting transition temperature $T_c^{onset}=0.84$ K at ambient pressure through temperature-dependent electrical resistivity measurements~\cite{wanggangBi2Ta3S6}. The Bi$_2$Ta$_3$S$_6$ layered structure consists of alternating 1H-TaS$_2$ transition-metal dichalcogenide (TMD) layers and honeycomb Bi layers, forming a natural quantum-well structure stacked along the $c$-axis. Previous works have found that 2H-TaS$_2$ is a superconductor with $T_c=0.8\ \mathrm{K}$~\cite{2024PhRvR...6c3218A,Tc-2HTMD-SR}.
Three questions are therefore naturally raised for Bi$_2$Ta$_3$S$_6$: (i) whether the superconductivity originates from the TaS$_2$ layers; (ii) whether the honeycomb Bi layers contribute nontrivial band topology near the Fermi level; and (iii) whether a TSC phase and MZMs can be realized.

In this work, using density functional theory (DFT) calculations, we report that the layered Bi$_{2}$Ta$_3$S$_6$-family superconductors exhibit nontrivial band topology with nonzero parity-based indices $(\mathbb{Z}_4;\mathbb{Z}_{2}\mathbb{Z}_{2}\mathbb{Z}_{2})=(2;000)$. Combined with the layer-resolved Wilson loop calculations, we find that the topological nature can be understood as the stacking of two QSHIs within a unit cell. Accordingly, we construct a 2D Bi-$p_{x,y}$ honeycomb model to capture the topological nature. The zigzag (ZZ) and armchair (AC) edge states are observed. Furthermore, the Fermi surface states are dominated by the TaS$_2$ layers, indicating that the superconductivity originates from the TaS$_2$ layers. In addition, our phonon calculations show that the series of Bi$_{2}$(Ta,Nb)$_3$(S,Se)$_6$ compounds (Bi$_{2}$Ta$_3$S$_6$-family) are all dynamically stable. Based on the electron-phonon coupling (EPC) and the McMilan formula, their superconducting transition temperatures ($T_c$) are estimated.
Moreover, using VASP2KP we obtain the Zeeman $g$ factors of the ZZ and AC edge states of Bi$_2$Ta$_3$S$_6$ from \emph{ab initio} calculations: $g_{x/y}^{\mathrm{ZZ}}=2.07/1.60$ and $g_{x/y}^{\mathrm{AC}}=0.50/0.06$. These results show strong anisotropy of the $g$ factors between the ZZ and AC edges.
Finally, we design a terrace structure of the topmost Bi monolayer on the (001) surface and find that the surface Bi layer becomes a second-order TSC under an magnetic field. Correspondingly, MZMs emerge at the corners of the Bi monolayer.
A similar paradigm also applies to the Bi$_{2}$Ta$_3$S$_6$ bulk with the emergence of Majorana hinge states.
Our results show that these natural quantum-well Bi$_2$Ta$_3$S$_6$-family superconductors offer ideal platforms for exploring TSC and MZMs.

\begin{figure}[t!]
\centering
\includegraphics[width=8.5 cm]{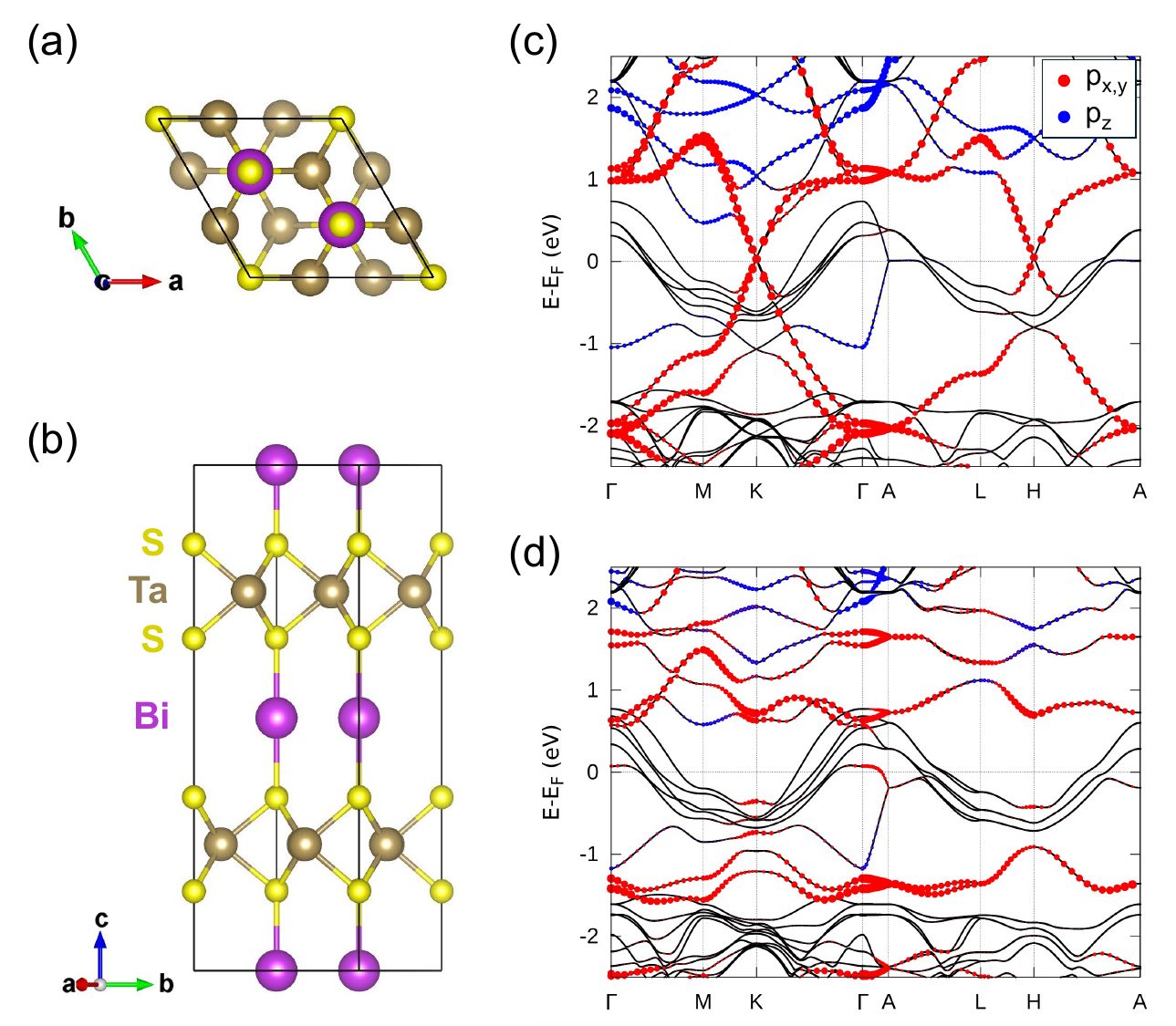}
\caption{
Crystal structure and electronic band structure of Bi$_{2}$Ta$_3$S$_6$.
(a,~b) Top and side views of the crystal structure. The formation of S–Bi–S bonds align the S atoms in the two TaS$_2$ layers.
(c) Electronic band structure  without SOC. The size of the red (blue) circles indicate the Bi-$p_{x,y}$ ($p_z$) orbital components. 
(d) Electronic band structure with SOC, where a nontrivial gap opens at K (H) of approximately 1.0 eV.
} \label{fig-BTS}
\end{figure}

\paragraph*{Crystal structure and band topology.---}
Bi$_{2}$Ta$_3$S$_6$ crystallizes in the centrosymmetric hexagonal space group $P6_3/mcm$. As illustrated in Figs.~\ref{fig-BTS}(a,b), the structure features triangular-prismatic TaS$_2$ layers arranged in a $\sqrt{3} \times \sqrt{3}$ supercell, while Bi atoms occupy positions directly above and below the S atoms with a 2/3 occupancy. The formation of S–Bi–S bonds aligns the S atoms in the two TaS$_2$ layers, in contrast to the Ta-atom alignment seen in the 2H-TaS$_2$ phase. These S–Bi–S bonds stabilize a flat, unbuckled honeycomb Bi lattice.

The PBE band structure is presented in Fig.~\ref{fig-BTS}(c), which exhibit a strong 2D feature. There is a pair of double Dirac cones at $K$ ($H$) originating from the two honeycomb Bi layers in a unit cell. Each Bi layer yields a Dirac cone at $K$ ($H$), formed by the $p_x$ and $p_y$ orbitals as shown in Fig.~\ref{fig-BTS}(c), while the Bi $p_z$ orbital states lie far from the Fermi level ($E_F$) due to the Bi-S bonding.
When spin-orbit coupling (SOC) is included [Fig.~\ref{fig-BTS}(d)], the double Dirac cones open a nontrivial gap. Due to the strong SOC, the nontrivial gap at $K/H$ is about $1.0$ eV. In addition, since the Dirac-cone states of the Bi layers are gapped out by SOC, the Fermi-surface states mostly originate from the floating bands of the TaS$_2$ layers, which contribute to the superconductivity~\cite{TaS2-Yang-Zhang}. It is worth noting that the slight band overlap near $\Gamma$ is removable, since the hybridization between the Ta-$d$ orbitals and the Bi-$p_{x,y}$ orbitals is negligible. Therefore, we compute the time-reversal invariant parity indices~\cite{four-Z2-prb,LC-Song,Z4-PRX,Z4-nc,Z4-prb,HOTI-MoTe2-Wang} $(\mathbb{Z}_4;\mathbb{Z}_{2}\mathbb{Z}_{2}\mathbb{Z}_{2})=(2;000)$, suggesting that the topological nature is regarded as the stacking of QSHI layers, \ie  the honeycomb Bi layers.

\begin{figure}[!t]
\centering
\includegraphics[width=9 cm, trim=5mm 5mm 0mm 0mm]{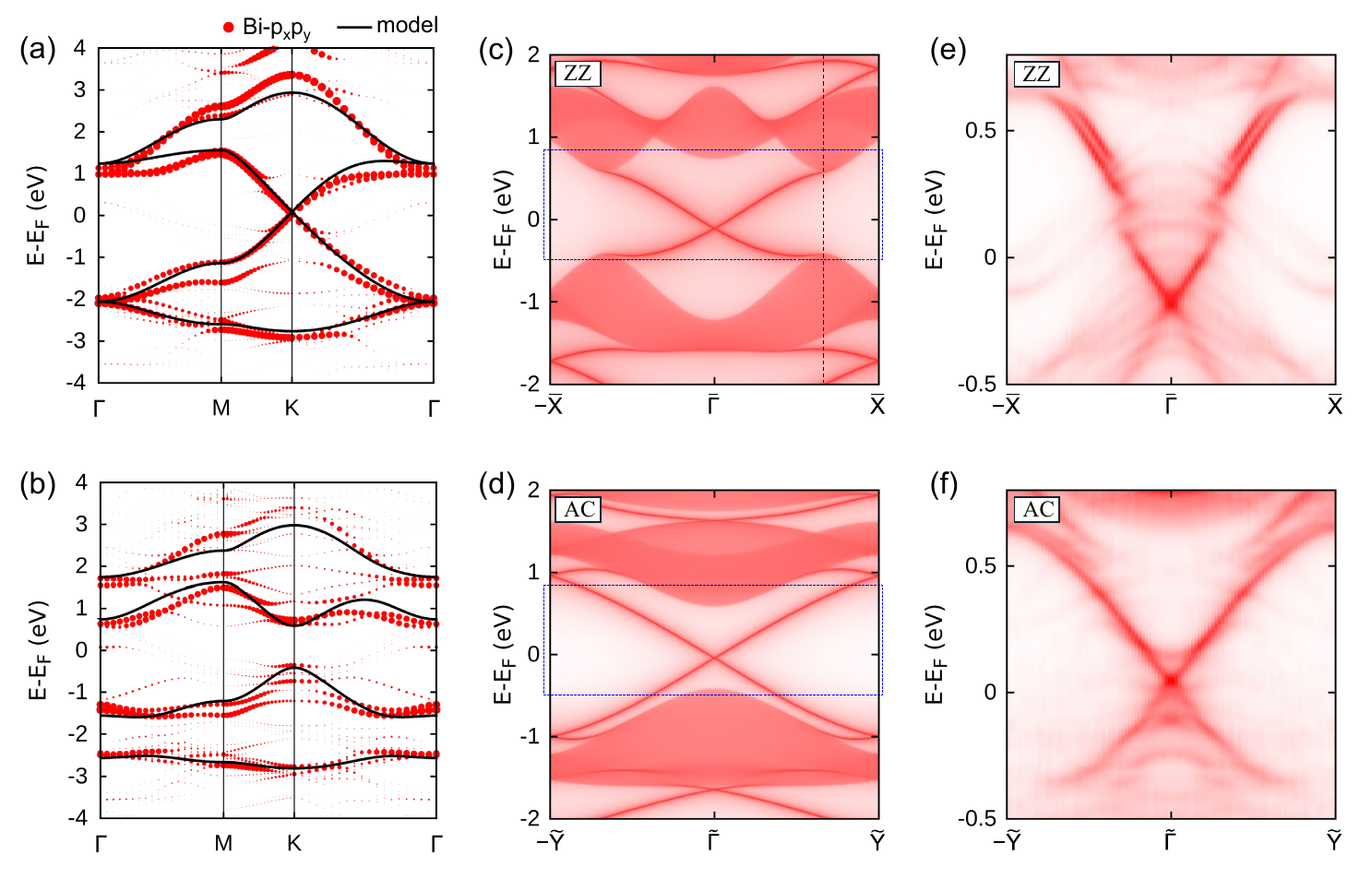}
\caption{
(a,~b) Electronic bands (black lines) of the Bi-$p_{x,y}$ honeycomb model Eq.~\pref{Bi-model} without and with SOC, which fit very well with the Bi-$p_{x,y}$ projections (red dots) from the DFT calculations.
(c,~d) Electronic bands of the ZZ-edge nanoribbon and the AC-edge nanoribbon with SOC, obtained from the Bi-$p_{x,y}$ honeycomb model.
(e,~f) DFT calculations of the ZZ-edge projections and AC-edge projections of the Bi layers in the Bi$_2$Ta$_3$S$_6$ slabs with SOC.
They agree well with those of our model (highlighted by the blue dashed boxes).
} \label{fig-model}
\end{figure}

\paragraph*{QSHI of the honeycomb Bi layer.---} 
The topological nature of the system is captured by a 2D honeycomb Bi-$p_{x,y}$ model~\cite{Bi-SiC},
\begin{equation}\label{Bi-model}
\begin{aligned}
    H(k)=\sigma_0\otimes H_0(k)+\lambda_{so}\sigma_z\otimes H_{so}.
\end{aligned}
\end{equation}
Here, $\sigma$ is the Pauli matrix for the spin degree of freedom. $H_0(k)$ includes nearest-neighbor and next-nearest-neighbor hopping of the $p_x$ and $p_y$ orbitals, while $H_{so}$ is the onsite SOC term. Both are specified in the Supplementary Material (SM). The parameters of the model are obtained by fitting the DFT band structure. In Figs.~\ref{fig-model}(a,b), the bands of the Bi-$p_{x,y}$ model closely match the fatted DFT bands. 
When SOC is included, a single pair of helical edge states emerges within the SOC-induced nontrivial band gap, confirming the QSHI nature of the system. 
The calculated edge states for the ZZ and AC edges of our model are presented in Fig.~\ref{fig-model}(c) and Fig.~\ref{fig-model}(d), respectively.
In addition, DFT calculations of Bi$_2$Ta$_3$S$_6$ in slab geometry clearly show the ZZ and AC edge states by side-surface projections in Figs.~\ref{fig-model}(e,f), which agree well with the model results.
These findings strongly suggest that the topological edge states can be experimentally detected on the side surfaces or terrace edges of the Bi$_{2}$Ta$_3$S$_6$ compounds.

\begin{figure}[!t]
\centering
\includegraphics[width=8.5 cm, trim=5mm 0mm 0mm 0mm]{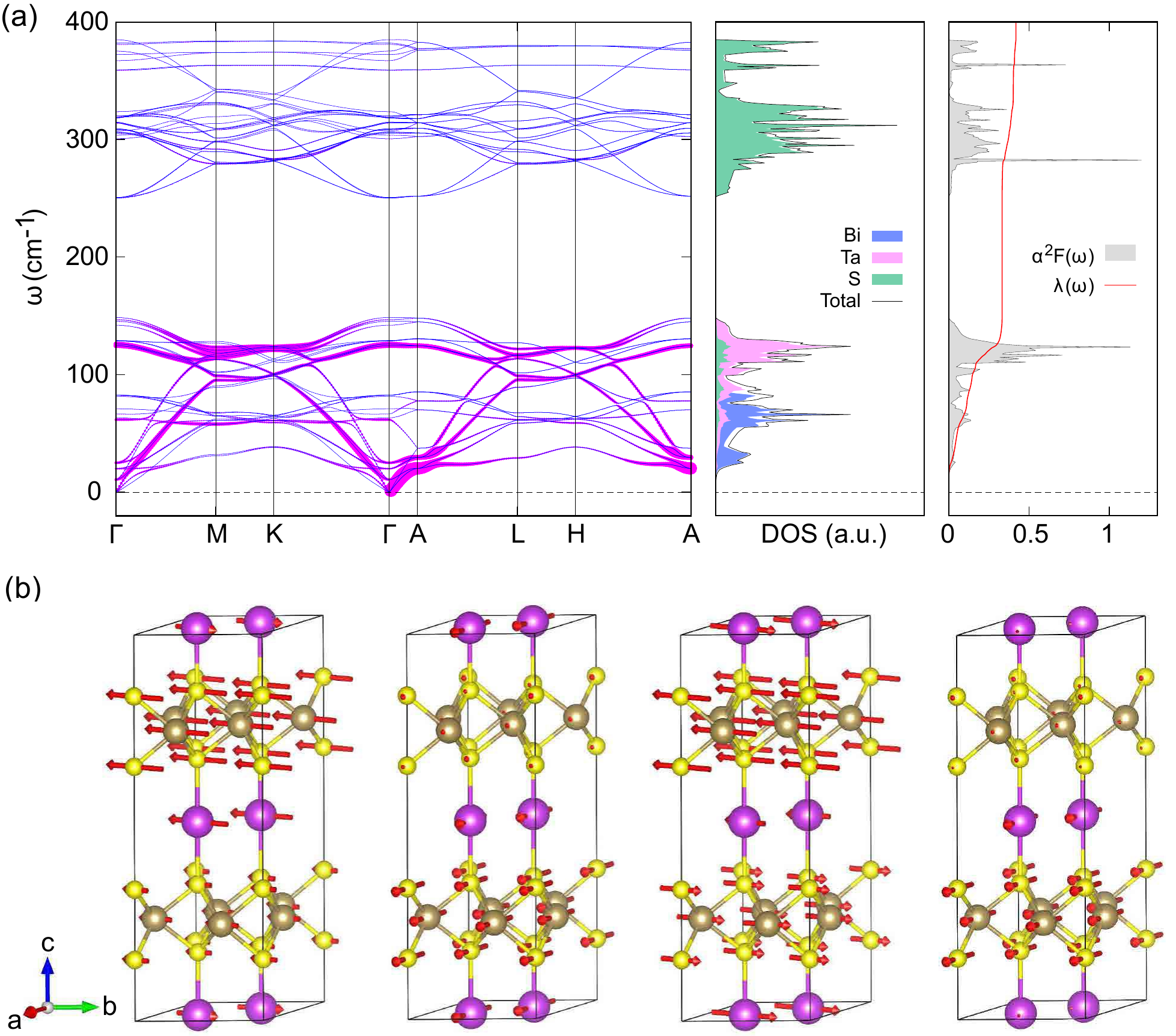}
\caption{
(a) Phonon properties and EPC for Bi$_2$Ta$_3$S$_6$.
Left panel: phonon dispersion obtained from DFPT calculations including SOC. The sizes of the fatted bands are proportional to the mode- and momentum-resolved EPC strength $\lambda_{q\nu}$.
Middle panel: total and atom-resolved phonon density of states (DOS). The low-frequency phonon modes are dominated by Ta and Bi atoms, while the high-frequency modes are dominated by S atoms.
Right panel: the Eliashberg spectral function $\alpha^2F(\omega)$ (grey-shaded area) and the cumulative coupling constant $\lambda(\omega)$ (red curve). Most of the EPC strength are contributed by the Ta phonon modes.
The effective screened Coulomb repulsion $\mu^* = 0.09$.
(b) Atomic vibration patterns of the four lowest-frequency phonon modes along $\Gamma$-A (near the A point).
} \label{fig3}
\end{figure}

\begin{figure}[t!]
\centering
\includegraphics[width=8.5 cm]{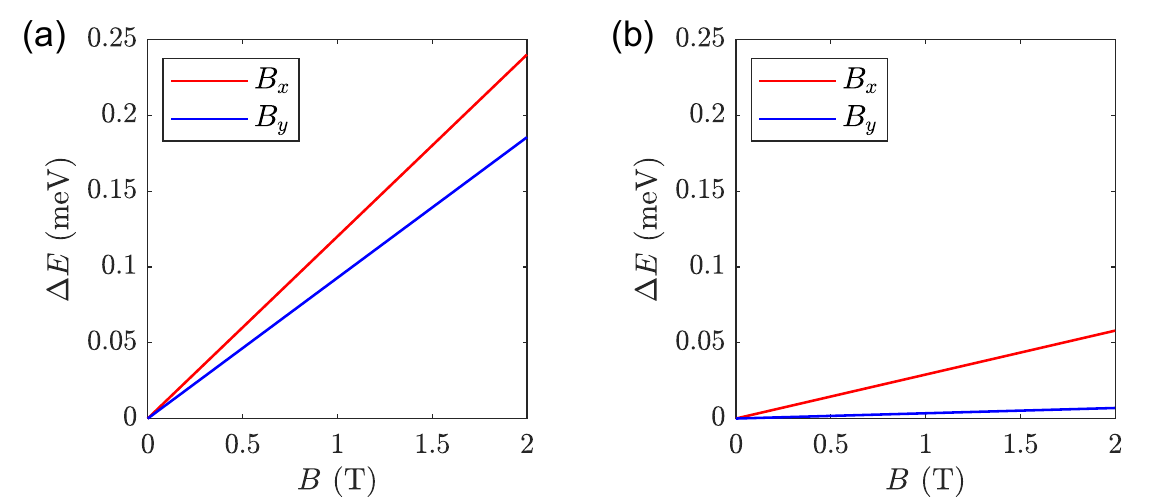}
\caption{
(a) Evolution of the Zeeman gap on the ZZ edge as a function of applied in-plane magnetic fields $B_x$ and $B_y$, obtained from DFT calculations and the VASP2KP package.
(b) Evolution of the Zeeman gap on the AC edge.
The effective $g$ factors of the ZZ (AC) edge can be extracted from the slopes: $g_x^\mathrm{ZZ}=2.07$ and $g_y^\mathrm{ZZ}=1.60$ ($g_x^\mathrm{AC}=0.50$ and $g_y^\mathrm{AC}=0.06$).
}
\label{fig-g factor}
\end{figure}

\begin{table}[!b]
    \begin{ruledtabular}
        \caption{Lattice parameters obtained from DFT optimization and superconducting transition temperatures $T_c$ for Bi$_2$Ta$_3$S$_6$-family layered compounds estimated using the Allen–Dynes modified McMillan equation Eq.~\pref{McMillan}.   These data of the 2H-TMD compounds are obtained in experiment~\cite{2024PhRvR...6c3218A}.}
        \begin{tabular}{lccc}
         Compound & $a$ (\AA) & $c$ (\AA) & $T_c$ (K) \\
        \hline
        Bi$_{2}$Ta$_{3}$S$_{6}$   & 5.763 & 17.581 & 0.81  \\
        Bi$_{2}$Ta$_{3}$Se$_{6}$  & 5.968 & 18.475 & 2.21 \\
        Bi$_{2}$Nb$_{3}$S$_{6}$   & 5.796 & 17.534 & 0.21 \\
        Bi$_{2}$Nb$_{3}$Se$_{6}$  & 5.994 & 18.431 & 1.60 \\
        2H-TaS$_2$        & 3.340    & 13.403     & 0.8 \\
        2H-TaSe$_2$       & 3.469    & 13.847     & 0.15 \\
        2H-NbS$_2$        & 3.345    & 12.921     & 5.5 \\
        2H-NbSe$_2$       & 3.470    & 13.630     & 7.3 \\
        \end{tabular}
        \label{table:tcoftmdc_single}
    \end{ruledtabular}
\end{table}

\paragraph*{Superconductivity in Bi$_{2}$Ta$_3$S$_6$ family.---} 
The dynamical stability of Bi$_{2}$Ta$_3$S${_6}$, Bi$_{2}$Ta$_3$Se$_6$, Bi$_{2}$Nb$_3$S$_6$, and Bi$_{2}$Nb$_3$Se$_6$ is confirmed by our phonon spectrum calculations, as evidenced by the absence of imaginary modes, as shown in the left panel of Fig.~\ref{fig3}(a) for Bi$_{2}$Ta$_3$S${_6}$ as an example (see others in the SM). The superconducting transition temperature $T_c$ is estimated using the Allen–Dynes modified McMillan equation:
\begin{equation}\label{McMillan}
T_c = \frac{\omega_{\mathrm{log}}}{1.2k_{B}} \exp\left[ \frac{-1.04(1+\lambda)}{\lambda(1 - 0.62\mu^{*}) - \mu^{*}} \right],
\end{equation}
where $k_B$ is the Boltzmann constant, $\mu^{*} $ is the effective screened Coulomb repulsion parameter, $\lambda$ is the EPC constant, and $\omega_{\mathrm{log}}$ is the logarithmic average phonon frequency. Using $\mu^{*} = 0.09$, the $T_c$ of Bi$_2$Ta$_3$S$_6$ is estimated to be 0.81 K, which is close to the experimental value. The calculated $T_c$ values are summarized in Table~\ref{table:tcoftmdc_single}.

Unlike the  2H-TaS$_2$ family—which hosts correlated electronic states such as charge density wave (CDW) and superconductivity, often in competition or coexistence~\cite{TaS2-Yang-Zhang,yujiabin2024,XxxNbSe2}—the intercalation of a honeycomb Bi monolayer in the Bi$_2$Ta$_3$S$_6$ family helps stabilize the structure and suppresses the CDW instability. The metallic bands of Bi$_2$Ta$_3$S$_6$ near $E_F$ originating from the TaS$_2$ layers were recently identified as unconventional metals in the obstructed atomic limit, where partially occupied elementary band representations (EBRs) reside at non-atomic positions near $E_F$~\cite{TaS2-Yang-Zhang}. These partially occupied EBRs contribute significant quantum geometry, which can lead to an enhancement of EPC~\cite{yujiabin2024}. In Fig.~\ref{fig3}(a), we show the mode- and momentum-resolved EPC $\lambda_{q\nu}$, the Eliashberg spectral function $\alpha^2F(\omega)$, and the frequency-accumulative coupling strength $\lambda(\omega)$, which together reveal that the dominant contribution to the total EPC stems from vibrations of the Ta sublattice and their associated $d$-orbital states. 
Fig.~\ref{fig3}(b) displays the vibrations of the four lowest-frequency modes carrying large $\lambda_{q\nu}$ along $\Gamma$–A near the A point, which are dominated by the Ta in-plane motions.
Moreover, the Bi monolayer introduces nontrivial topological states, making these materials natural quantum-well structures combining superconductors and QSHIs. As a result, these materials provide promising platforms for exploring topological superconductivity.



\begin{figure}[t!]
\centering
\includegraphics[width=8.2 cm, trim=5mm 0mm 0mm 0mm]{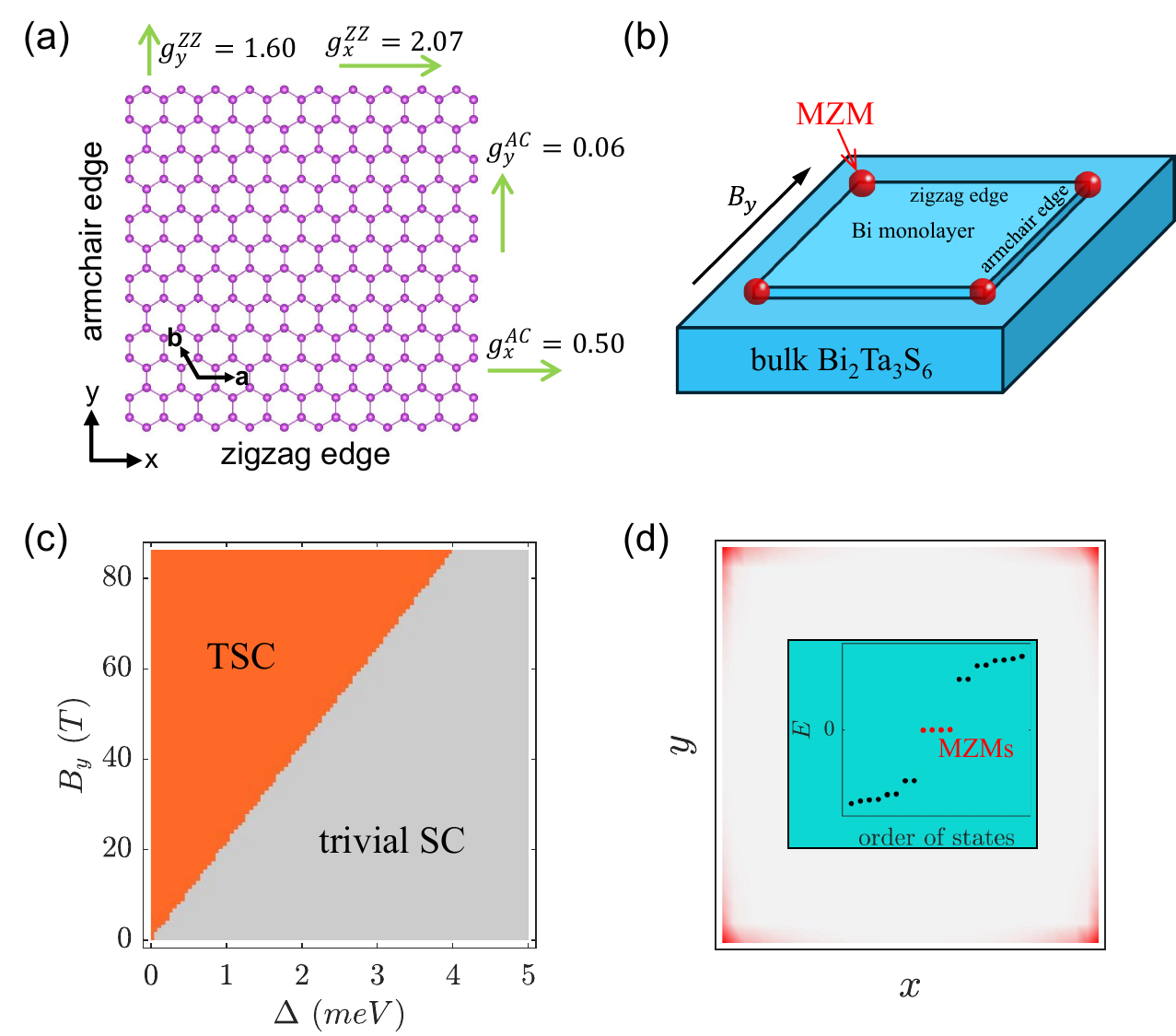}
\caption{
(a) A rectangle geometry of the honeycomb Bi monolayer with effective $g$ factors of the ZZ and AC edges obtained from DFT calculations.
(b) Schematic plot of corner MZMs realized in Bi$_2$Ta$_3$S$_6$ superconductor.
(c) Phase diagram for a $y$-directed magnetic field $B_y$ versus superconducting pairing $\Delta$, with $\alpha=1.85$ and $\mu=-106.7$ meV (which sets the ZZ-edge Dirac point at $E_F$). The orange region denotes the TSC phase that hosts corner MZMs. 
(d) Probability densities of the four MZMs (colored red) in a square geometry, which are localized at the four corners between the ZZ edges ($x$-directed) and the AC edges ($y$-directed). Inset: energy levels showing four MZMs. The parameters with $\Delta=0.123$~meV and $ B_y=4.24$~T reside in the TSC phase and host corner MZMs.
}
\label{fig-MZM}
\end{figure}

\paragraph*{The $g$ factors of topological edge states.---}
To compute the $g$ factors of the edge states, we first perform nanoribbon calculations with the Bi layer placed in the $xy$ plane, imposing an open boundary condition along $y$ and a periodic boundary condition along $x$ (without loss of generality). We then construct the edge $k\!\cdot\!p$ Hamiltonian in the absence of a magnetic field~\cite{Zhang-Sheng-cpl}, $\hat{H}^{kp}(k_x)=\hat{H}_0+\frac{\hbar^2 k_x^{2}}{2m_e}+\frac{\hbar}{m_e}k_x\,\hat{\pi}_x$, where the generalized momentum operator is $\hat{\boldsymbol{\pi}}=\hat{\boldsymbol{p}}+\frac{1}{2m_e c^2}\bigl[\hat{\boldsymbol{s}}\times\nabla V(\boldsymbol{r})\bigr]$, with $\hat{\boldsymbol{p}}=-\mathrm{i}\hbar\nabla$ the canonical momentum, $V(\boldsymbol{r})$ the crystal potential, and $\hat{\boldsymbol{s}}$ the spin operator. Further considering an in-plane magnetic field, the responses to $B_x$ and $B_y$ can be markedly different as follows,
\begin{equation}
\begin{aligned}
    &\begin{cases} \bB=B_y\hat{\be}_y,\ \bA=(B_y z,0,0), \\ \hat{H}(B_y,k_x)=\hat{H}^{kp}\!\left(k_x+\frac{eB_y\hat{z}}{\hbar}\right)+\frac{2\mu_B}{\hbar}B_y\hat{s}_y,
    \end{cases}
    \\
    &\begin{cases} \bB=B_x\hat{\be}_x,\ \bA=(0,-B_x z,0), \\ \hat{H}(B_x,k_x)=\hat{H}^{kp}(k_x)+\frac{(eB_x\hat{z})^2}{2m_e}-\frac{eB_x\hat{z}}{m_e}\hat{\pi}_y+\frac{2\mu_B}{\hbar}B_x\hat{s}_x.
    \end{cases}
\end{aligned}
\end{equation}
Here, $\mu_B$ is the Bohr magneton and $\hat{z}$ is the position operator along $z$. In Figs.~\ref{fig-g factor}(a,b), we show the band spitting $\Delta E$ at the $\Gamma$ point as a function of the in-plane magnetic field $\bB$ for the ZZ and AC edge, respectively. One can extract the effective $g$ factors via $\Delta E=g\mu_B B$: $g_x^\mathrm{ZZ}=2.07$ and $g_y^\mathrm{ZZ}=1.60$ ($g_x^\mathrm{AC}=0.50$ and $g_y^\mathrm{AC}=0.06$) for the ZZ (AC) edge~\cite{gfactor-zs}.

\paragraph*{Majorana zero modes.---}
To exploit the coexistence of the QSHI layer and intrinsic superconductivity of the Bi$_{2}$Ta$_3$S$_6$-family compounds, we consider a terrace structure with
the Bi-terminated (001) surface to host MZMs at corners between the ZZ and AC edges, as shown in Fig.~\ref{fig-MZM}(b).
The honeycomb Bi monolayer on the surface can be obtained either by exfoliating the bulk Bi$_{2}$Ta$_3$S$_6$-family superconductors or by epitaxially growing a Bi monolayer on 2H-phase TaS$_2$-family superconductors.
Under the superconducting proximity effect, the superconductivity induced on the topological Bi-honeycomb layer can be described in Bogoliubov–de Gennes (BdG) representation, 
\begin{equation}\label{Bi-BdG}
\begin{aligned}
    H_{\text{BdG}}(\bk)=\left(\begin{array}{cc}
    H(\bk)+H_B-\mu & -i\Delta\sigma_y \\
    i\Delta\sigma_y & -H^*(-\bk)-H_B+\mu
    \end{array}\right).
\end{aligned}
\end{equation}
Here, $H_B=\frac{1}{2}\alpha\mu_BB_y\sigma_y$ is the in-plane Zeeman term. $\Delta$ is the superconducting pairing potential and $\alpha$ is the $g$ factor of the model for the $y$-directed Zeeman field. In our model, the induced $g_y$ factors for the ZZ and AC edge states are $1.6$ and $0.0$, which are quantitatively close to the DFT results [Fig.~\ref{fig-MZM}(a)]. In addition, the effective Zeeman field on the edges can be introduced by either an external magnetic field or a ferromagnetic attachment. In this scenario, the Zeeman gap competes with the superconducting gap on the topological edge states.

The key to generating corner MZMs is to fabricate corners between magnetism-dominated edges and superconductivity-dominated edges~\cite{gapcompete,cornerMZM-prl,cornerMZM-prb}.
On the AC edge, the gap is induced by the superconducting pairing $\Delta$. On the ZZ edge, as the Zeeman field increases, the edge gap closes and then reopens, becoming dominated by the magnetism.
This gap closure indicates a topological phase transition and the topmost Bi layer becomes a second-order TSC after the ZZ edge is dominated by magnetism. Therefore, we determine the phase diagram against $B_y$ and $\Delta$ in Fig.~\ref{fig-MZM}(c) by the gap closure on the edge states, where the TSC region is denoted in orange.
Beyond a critical value, the in-plane Zeeman field induces opposite effective Dirac masses between the ZZ and AC edges, leading to corner MZMs.
A similar paradigm also applies to the bulk Bi$_{2}$Ta$_3$S$_6$ family with the emergence of Majorana hinge states.

\paragraph*{Conclusion and discussion.---}
In summary, we have uncovered nontrivial band topology in the Bi$_{2}$Ta$_3$S$_6$-family superconductors, which combine topological Bi-honeycomb QSHI layers with superconductivity derived from TaS$_2$ layers, making them natural quantum-well platforms for exploring topological superconductivity.
As a result of the band topology, the ZZ and AC edge states of the Bi monolayer are obtained.
Using VASP2KP, we find a strong anisotropy of effective $g$ factors between the ZZ and AC edge states, with $g_{x/y}^{\mathrm{ZZ}}=2.07/1.60$ and $g_{x/y}^{\mathrm{AC}}=0.50/0.06$. 
Benefiting from such an strong anisotropy of the edge-state $g$ factors, we design a terrace structure of the topmost Bi monolayer on the (001) surface under a uniform external magnetic field, to realize a TSC phase. We therefore find that MZMs emerge at corners between the ZZ and AC edges when they have opposite effective Dirac masses. In practice, the external magnetic field can be replaced by magnetic attachments to the outermost Bi ZZ edges.
Furthermore, using phonon spectrum calculations, we have verified that a Bi-honeycomb layer grown on superconducting 2H-TaS$_2$-family substrates is dynamical stable, which provides another class of platforms for realizing TSC and MZMs.
Finally, we suggest that the edge states of the outermost Bi layer are accessible to angle-resolved photoemission spectroscopy, and that the MZMs in the TSC phase are detectable as zero-bias conductance peaks in low-temperature scanning tunneling microscopy measurements.

\ \\
\noindent{\bf ACKNOWLEDGEMENTS} \\
This work was supported by National Key R\&D Program of China (Grants No. 2023YFA1607400, No. 2022YFA1403800, and No. 2022YFA1403900), the National Natural Science Foundation of China (Grants No. 12188101, No. 12504046, and No. 52325201),  and the Center for Materials Genome.  \\

\noindent{\bf REFERENCES} \\

\bibliography{TSC}


\clearpage

\end{document}